\documentclass[conference]{IEEEtran}
\usepackage{array}
\usepackage{booktabs}
\usepackage{listings}

\usepackage[shortlabels, inline]{enumitem}
\usepackage{graphicx}
\usepackage[table]{xcolor}
\usepackage[numbers]{natbib}

\usepackage[hidelinks]{hyperref}
\usepackage[final]{microtype}
\usepackage{fancyvrb}
\usepackage{tabu}
\usepackage{booktabs}
\usepackage{todonotes}
\usepackage{marginnote} 

\usepackage{cleveref}
\usepackage{amsmath}
\makeatletter
\def\@copyrightspace{\relax}
\makeatother
\usepackage{flushend}
\def\BibTeX{{\rm B\kern-.05em{\sc i\kern-.025em b}\kern-.08em
        T\kern-.1667em\lower.7ex\hbox{E}\kern-.125emX}}

\crefname{table}{tab.}{tab.}
\crefname{figure}{fig.}{fig.}

\begin{document}
\title{An Exploratory Survey of Hybrid Testing Techniques Involving Symbolic Execution and Fuzzing}

\author{\IEEEauthorblockN{Saahil Ognawala, Ana Petrovska, Kristian Beckers}
    \IEEEauthorblockA{\textit{Technical University of Munich, Germany} \\
        \{ognawala,petrovsk,beckersk\}@in.tum.de}
}

\maketitle
\definecolor{orange}{rgb}{1,0.5,0}



\begin{abstract}
Recent efforts in practical symbolic execution have successfully mitigated the path-explosion problem to some extent with search-based heuristics and compositional approaches. Similarly, due to an increase in the performance of cheap multi-core commodity computers, fuzzing as a viable method of random mutation-based testing has also seen promise. However, 
the possibility of \emph{combining symbolic execution and fuzzing}, thereby providing an opportunity to mitigate drawbacks in each other, has not been sufficiently explored. Fuzzing could, for example, expedite path-exploration in symbolic execution, and symbolic execution could make seed input generation in fuzzing more efficient. There have only been, in our view, very few hybrid solution proposals with symbolic execution and fuzzing at their centre. By analyzing \emph{77 relevant and systematically selected papers}, we
\begin{enumerate*}[(1)]
	\item present an overview of hybrid solution proposals of symbolic execution and fuzzing,
	\item perform a gap analysis in research of hybrid techniques to improve \emph{both}, plain symbolic execution and fuzzing, 
	\item propose new ideas for hybrid test-case generation techniques.
\end{enumerate*} 
\end{abstract}
\section{Introduction}\label{sec:introduction}
Symbolic execution \cite{king1976symbolic} and fuzzing \cite{miller1990empirical,oehlert2005violating,sutton2007fuzzing} are well-known methods of software testing, which have been addressed over the past couple of decades as techniques for generating test-cases and finding bugs in software. Symbolic execution is a technique to deterministically analyze the system-under-test (SUT) by enumerating the program paths as constraint systems that are solved (using constraint solvers) to generate test cases that execute respective paths. A program path is defined as a directed sequence of branching conditions starting from any entry point of the program (e.g.\ \emph{main} or an open API function call) to any exit point (e.g.\ \emph{return} or \emph{throw} statement). Some practical approaches to symbolic execution are concolic execution \cite{cadar2005execution,sen2005cute}, whitebox fuzzing \cite{godefroid2008automated} and bounded model checking (BMC) \cite{clarke2003behavioral}. Concolic (\textbf{conc}rete + symb\textbf{olic}) execution is a technique where the constraints along all paths of a program are gathered by a concrete execution of the program. Whitebox fuzzing \cite{godefroid2008automated} is a practical approach to symbolic execution that uses seed inputs, like blackbox fuzzing, to record some initial constraints of a program which are sequentially negated to form new constraints and solved to generate new test cases. BMC is a similar method for \emph{model checking} using constraint solvers, with some optimizations such as \emph{loop-unrolling}\footnote{Please note that due to this commonality in the underlying concept, we will treat the terms, \emph{concolic execution}, \emph{whitebox fuzzing} and \emph{BMC}, as synonyms for \emph{symbolic execution}, as explained in \cref{sec:classification}.}. Symbolic execution suffers from two main problems in practical application. First of these is the path-explosion problem, which means that for any program that contains non-trivial structures, such as input-dependant loops, recursion or a highly compositional architecture, the number of paths that need to be explored by the symbolic execution engine \emph{may be} infinite or very large. In such cases, symbolic execution needs to be stopped before it may have explored all interesting paths in a program. The second problem with symbolic execution is its reliance on constraint solvers, such as SMT and Z3, which often take a long time to prove satisfiability or return counter-examples for large constraint systems.  

Fuzzing (also called \emph{blackbox fuzzing}) \cite{sutton2007fuzzing} is an automated technique of random testing, with monitoring mechanisms to perform \emph{mutations} on \emph{seed inputs}, so as to uncover new functionality. \emph{Mutation} simply means changing some bits in an input buffer to generate a new input value. \emph{Seed inputs} are the initial inputs that are (generally) manually chosen by the tester. Some modern fuzzing tools, such as AFL \cite{lcamtuf2016afl}, may even use genetic algorithms and code instrumentation for tracking explored program paths and mutating the seed inputs. The main problem, however, associated with na\"{i}ve fuzzing is that, due to its relative ``blindness'' to the internals of a program, it misses many program paths that have branching conditions that are hard to satisfy, thereby leading to low path-coverage. 

Due to the similarity in their goals, symbolic execution and fuzzing have the potential to be used as mutually beneficial methods for discovering vulnerabilities. Within reasonable time limits, fuzzing is an effective technique for exploring \emph{some} paths in a program in full depth. Symbolic execution, on the other hand, is an effective technique for exploring most branches in a program at low-depths (closer to an entry point). 
An example of a trivial 2-step hybrid strategy involving symbolic execution and fuzzing is as follows:
\begin{enumerate*}[(1)]
	\item Fuzz the program for a short amount of time to explore the most easy-to-reach paths. 
	\item Target those branches with symbolic execution that were not taken by fuzzing. 
\end{enumerate*}
In the above example, we have proposed to use fuzzing with its original \emph{input mutation} strategy, and a modification in the \emph{path search} strategy of symbolic execution. There are other \emph{technical aspects} of symbolic execution and fuzzing, similar to path search and input mutation, that may be used to increase the effectiveness of these two techniques. We would like to organize the field of hybrid techniques of fuzzing and symbolic execution (henceforth referred to as, simply, \emph{hybrid techniques} or \emph{hybrid solutions}) as viewed from the perspective of such technical aspects. 

\subsection{Research Questions}\label{sec:research-questions}
To describe state-of-the-art in hybrid symbolic execution and fuzzing methods, we answer the following \emph{three} research questions (RQs) in this paper
\begin{enumerate}	
	\item{\emph{RQ1: What is the publication trend in the field of fuzzing and symbolic execution?}} We determine if the field of hybrid techniques is growing, by analyzing the trend of publications.
	
	\item{\emph{RQ2: Which novel hybrid solutions have been proposed over the years?}} We consider solutions proposals that include symbolic execution and fuzzing in any individual capacity. The rationale behind this question is to determine how many publications have proposed truly hybrid techniques utilizing both participating methods in a demonstrable manner. 
    
	\item{\emph{RQ3: Which technical aspects of symbolic execution and fuzzing have been utilized in the novel solution proposals?}} For novel solution proposals, we determine the technical aspects of symbolic execution and fuzzing which have been addressed or utilized in them. We ask this to perform a gap analysis in research of hybrid techniques. 
\end{enumerate}
\subsection{Related Work}\label{sec:background}
To the best of our knowledge, hybrid solutions involving symbolic execution and fuzzing have never been analyzed empirically before.
In a survey in 1999 \citeauthor{edvardsson1999survey} \cite{edvardsson1999survey} covered symbolic execution as a technique for automatic test case generation. Other surveys and meta-studies \cite{puasuareanu2009survey,cadar2011symbolic,cadar2013symbolic}, that have dealt exclusively with practical symbolic execution, evaluation results and overview of optimizations in symbolic execution, have also been published in the past. 
Fuzzing was comprehensively discussed as a tool for finding vulnerabilities by \citeauthor{sutton2007fuzzing} \cite{sutton2007fuzzing}. \citeauthor{van2005fuzzing} presented an overview of the research in the field of fuzzing in \cite{van2005fuzzing}, as did \citeauthor{takanen2009fuzzing} in \cite{takanen2009fuzzing}. 
Symbolic execution and fuzzing have also been considered as vulnerabilities discovery tools in the single works \cite{zhang2015survey,li2015survey}. 
Some of the papers analyzed by us, including the ones listed above, were classified (\cref{sec:classification}) as \emph{state-of-the-art} papers, but we found no papers that described the state-of-the-art in hybrid studies of symbolic execution and fuzzing. 
We hope that this empirical study, and the implications thereof, as provided by us, would prove to be ultimately useful in conducting effective research in the field of hybrid techniques.

The rest of the paper is divided as follows -- In \cref{sec:method} we describe the methodology of our research, including the data collection procedure and systematic classification by voting. In \cref{sec:results}, we use the results of our classification to answer the main research questions. This is followed by a broader discussion of the results and a summary of the state-of-the-art in \cref{sec:interpretation}. We explicitly give ideas for future work in \cref{sec:visions}. Finally, we conclude the paper in \cref{sec:conclusion}.  
\section{Methodology}\label{sec:method}
\begin{figure}
    \centering
    \includegraphics[width=0.9\linewidth, keepaspectratio=true]{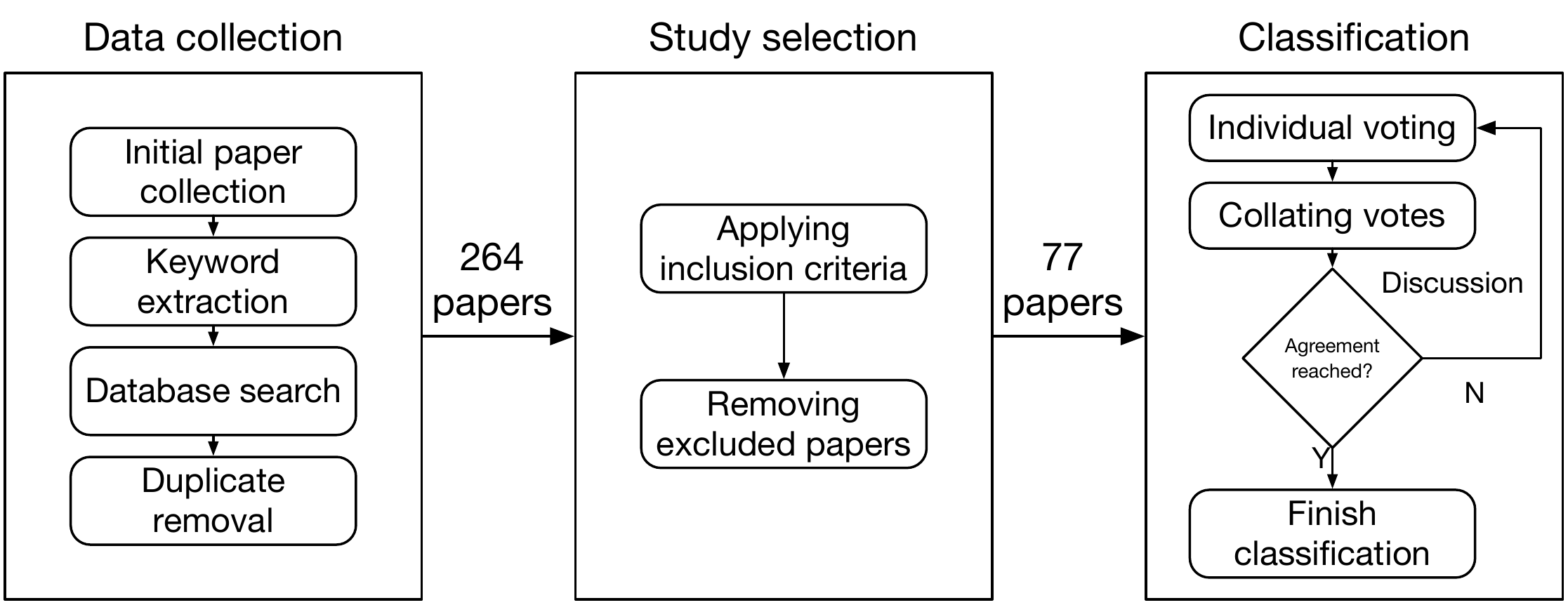}
    \caption{Overview .of the methodology}
    \label{fig:analysis-overview}
\end{figure}

\vspace{-1ex}
Our study was designed by following guidelines \cite{kitchenham2009systematic} for a systematic mapping study to obtain and organize an exhaustive list of publications. This was combined with the standard procedures of a systematic literature review prescribed by \citeauthor{keele2007guidelines} \cite{keele2007guidelines}, which lead to an in-depth analysis of the publications. An overview of the analysis process related to the obtained publications is depicted in \cref{fig:analysis-overview}. 

We, first, describe the systematic procedure for data collection and study selection. Then, in order to answer our research questions, we define the classification criteria which we applied to our final dataset of included publications. There may be some internal and external threats to validity in our own study, eg.\ our search strings may have led to exclusion of certain papers whose abstracts do not explicitly mention our search keywords, but are nevertheless relevant. Please note that all papers in our dataset were published no later than 2016. 

We have made the full dataset available through our online repository \cite{datarepo} (\emph{identifying information removed for double-blind review}). This dataset includes raw search results, papers passing inclusion criteria, individual votes from authors and the consolidated decisions. 

\subsection{Data Collection} \label{sec:data_collection}
We started the data collection procedure with a search on popular bibliographic databases. Search keywords were primarily based on an \emph{initial set} (size 21) of papers. We included some of these papers based on our domain knowledge, knowing them as key contributions in the field of fuzzing and symbolic execution. The rest of the initial papers were obtained by \emph{snowballing} through the related work of the initial set, as described by \citeauthor{wohlin2014guidelines} \cite{wohlin2014guidelines}. Primary search keywords were derived from the initial and snowballed sets by creating a word-cloud on the combined texts of their abstracts and choosing the most commonly occurring (stemmed) words. The main keywords derived, therefore, were \{``\texttt{test}'',\ ``\texttt{symbol}'',\ ``\texttt{execute}'',\ ``\texttt{fuzz}''\}. These search keywords were modified, as shown in \cite{datarepo}, to fit the \emph{advanced search} syntax for all chosen databases. In an initial exploration phase, we also used keywords such as ``\texttt{random}'' and ``\texttt{random test OR fuzz}'', because fuzzing is essentially a more systematic way of random testing, but observed that the search results grew very large with many irrelevant papers showing up at the top of the results. 

The databases chosen to perform the search were
\begin{enumerate*}[(1)]
	\item ACM Digital Library (\emph{10 papers}),
	\item IEEE Xplore (\emph{18 papers}),
	\item Scopus (\emph{31 papers}),
	\item SpringerLink (\emph{212 papers}), and
	\item CiteSeerX (\emph{9 papers}).
\end{enumerate*}

The next step in the collection process was to simply remove the duplicate results. These results could be 
\begin{enumerate*}[1)]
    \item same paper appearing in more than one databases, or
    \item different versions of the same paper e.g.\ extended journal version of a conference proceedings paper. 
\end{enumerate*}

After removing duplicates from results from all five search engines, we were left with \emph{264 unique publications} matching the search keywords. 
\subsection{Study Selection}
To ensure that the papers used for classification were from software engineering field and, at least, contributed to symbolic execution \emph{or} fuzzing, we needed to apply the following \emph{inclusion criteria} \cite{keele2007guidelines} to each of them --
\begin{enumerate}
\item Is the paper related to software testing or engineering?
\item Does the paper contribute to symbolic execution and/or fuzzing? 
\end{enumerate}
These inclusion criteria were manually applied by the first author, and only those papers were selected for classification for which the answers to both of the above questions were \emph{``yes''}. 

After applying the inclusion criteria, there remained \emph{77 selected papers} that we used for the classification stage and for answering all relevant research questions.
\subsection{Classification}\label{sec:classification}
We classified all included papers according to \emph{six criteria}. The final classification was decided by a full majority in a three-way voting by all authors. This classification would be useful in answering research questions, as listed in \cref{sec:research-questions}. The categories for all classification criteria were obtained by analyzing the key terms and concepts addressed in our initial set. For most papers, we were able to assign categories in all criteria by looking at the abstract alone. However, for some papers, especially the ones without a clear majority, we read the full paper texts to decide upon a category. The classification criteria are as follows. 

\subsubsection{Criterion 1: General Field of Contribution}
The choices for this classification criterion were 
\begin{enumerate}[a.]
    \item Symbolic execution,
    \item Fuzzing, and
    \item Both
\end{enumerate}
The first classification criterion, if applied correctly, also served as a validation for the inclusion criteria because if a paper can not, effectively, be classified as one of the above three choices, then it should have been excluded from the study space in the first place. 

\subsubsection{Criterion 2: Introduction of a Novel Technique}
This criterion was to identify the field where a new solution has been proposed. The choices for this criterion were
\begin{enumerate}[a.]
    \item Symbolic execution,
    \item Fuzzing, 
    \item Hybrid technique, and
    \item None
\end{enumerate}
The hybrid category was chosen only if, 
\begin{enumerate*}[(i)]
	\item the solution utilizes both, fuzzing and symbolic execution, or
	\item solution involves the use of both fuzzing and symbolic execution, but it is not obvious which one of these two techniques have been primarily subjected to a modification. 
\end{enumerate*}
We selected ``None'' when the paper in question does not propose a new solution, but, as we describe in the following criteria, contributed in a different manner. 

\subsubsection{Criterion 3: Description of State-of-the-art}
This criterion was to find if the paper presents a systematic state-of-the-art study or any meta-study, like our own. The choices for answering this questions were, also,
\begin{enumerate}[a.]
    \item Symbolic execution,
    \item Fuzzing, 
    \item Hybrid technique, and
    \item None.
\end{enumerate}

\subsubsection{Criterion 4: Novelty in the Solution for Fuzzing} 
This criterion enumerates the technical aspects of fuzzing. The choices for this criterion were
\begin{enumerate}[a.]
\item{Input mutation} -- includes modification in mutation strategy of random fuzzing, 
\item{Static analysis} -- uses a \emph{code analysis} technique such as static analysis, information flow analysis or symbolic execution to optimize fuzzing, 
\item{Expert guidance} -- uses any technique, other than code analysis, to optimize fuzzing, and 
\item{No modification} -- does not propose optimizing any technical aspects of fuzzing.
\end{enumerate}
This list of fuzzing aspects was determined by the authors using state-of-the-art descriptions in \cite{sutton2007fuzzing,lcamtuf2016afl}. 

\subsubsection{Criterion 5: Novelty in the Solution for Symbolic Execution}
This criterion enumerates the technical aspects of symbolic execution. The choices for this criterion were
\begin{enumerate}[a.]
\item{Path search} -- proposes a new path search strategy, such as directed search, or modification of an old path search strategy, 
\item{Compositional analysis} -- proposes to treat the SUT \emph{compositionally}. This means disintegrating a modular system and analyzing interactions of individual components, 
\item{Constraint solving} -- proposes an optimization in the constraint solver of the symbolic execution engine, and 
\item{No modification} -- does not propose optimizing any technical aspects of symbolic execution. 
\end{enumerate}
The list of symbolic execution aspects was determined by the authors using the state-of-the-art description in \cite{cadar2013symbolic} and recent solution proposals such as \cite{christakis2015ic,ma2011directed}. 

\subsubsection{Criterion 6: Description of Novel Target-system in the Paper}
In this criteria, we tried to find if the paper provides a new implementation or evaluation of an existing symbolic execution or fuzzing solution. The choices for answering this questions were -- 
\begin{enumerate}[a.]
    \item Yes, and
    \item No.
\end{enumerate} \vspace{4ex}
\section{Results}\label{sec:results}
\subsection{RQ1: Publication Trend}\label{sec:result-rq1}
To answer this question, we look at classification criteria \emph{1}, \emph{2}, \emph{3} and \emph{6}. The time-wise trend of publications in hybrid techniques is described as a combination of novel solution proposals and other studies. These other studies may be state-of-the-art studies, such as \cite{van2005fuzzing,takanen2009fuzzing}, or papers that simply introduce new target implementations of existing techniques, such as \cite{chen2009ewap}.
\begin{table}[]
    \centering
    \fontsize{8}{10}\selectfont
    \caption{Classification of papers obtained with systematic search (Not mutually exclusive)}
    \label{tab:all-papers}
    \begin{tabu}{X X r}
        \toprule
        \multicolumn{1}{l|}{\textbf{Category}}                           & \multicolumn{1}{l|}{\textbf{Sub-category}} & \textbf{Number} \\ \midrule
        All publications                                        &                       & 77     \\ \midrule
        Novel solutions                                         &                       & 65     \\ 
        \rowfont{\color{gray!120}}& Symbolic execution                 & 43     \\ 
        \rowfont{\color{gray!120}} & Fuzzing               & 13     \\ 
        \rowfont{\color{gray!120}} & Hybrid                & 9      \\ 
        \hline
        State-of-the-art studies                                &                       & 9      \\ \midrule
        New target implementations                              &                       & 24     \\ \midrule \midrule
        Novel solutions and state-of-the-art studies            &                       & 1      \\ \midrule
        Novel solutions and new target implementations          &                       & 20     \\ \midrule
        State-of-the-art studies and new target implementations &                       & 0      \\ \midrule
    \end{tabu}
\end{table}
As shown in \cref{tab:all-papers}, most of the results returned by our search strings on the databases were novel solution proposals (65). These solutions proposals are in the field of symbolic execution, fuzzing or hybrid methods. Other than the novel solution proposals, there are state-of-the-art studies (9) and implementation reports for novel target systems (24). As we can see from \cref{tab:all-papers}, we also found that many of the publications (20) propose novel solutions and, in the same paper, describe implementations for new target systems. We only found 1 paper in our analysis that proposes a novel solution and a state-of-the-art study, while no paper presented a state-of-the-art study and new target implementation together. 
\begin{figure}[h]
	\centering
	\includegraphics[width=0.9\linewidth,keepaspectratio=true]{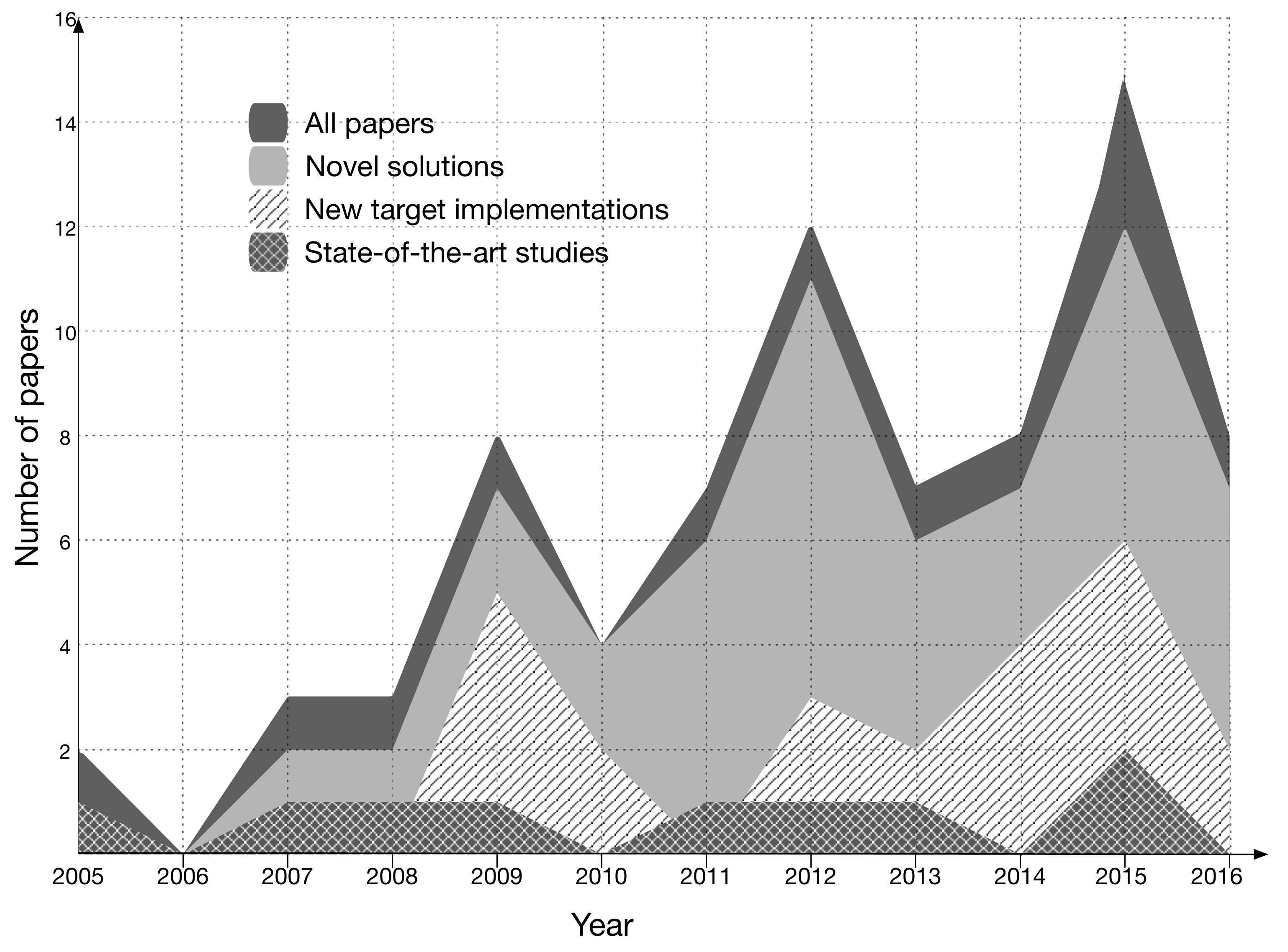}
	\caption{Number of publications by year}
	\label{fig:time-trend}
\end{figure}

\paragraph*{\textbf{Interpretation}} 
Looking at the time-wise distribution of publications, as shown in \cref{fig:time-trend}, the earliest obtained papers matching our inclusion criteria are from 2005 and solution proposals have been increasing since then, albeit with a few slump years in between (e.g.\ no papers in 2006). As expected, the number of state-of-the-art studies and implementations for new targets only appear after some seminal novel solutions, such as \cite{van2005fuzzing,cadar2005execution}, were published that proposed feasible solutions for modern platforms. However, as we will see in the next two research questions, these solutions did not always involve, both, symbolic execution and fuzzing.

The interpretations derived from the results of RQ1 may be trivial, but help us in deriving more insights about the results of other research questions. 

\subsection{RQ2: Hybrid Solutions}\label{sec:result-rq2}
\begin{figure}[h]
    \centering
    \includegraphics[width=\linewidth, height=0.5\linewidth]{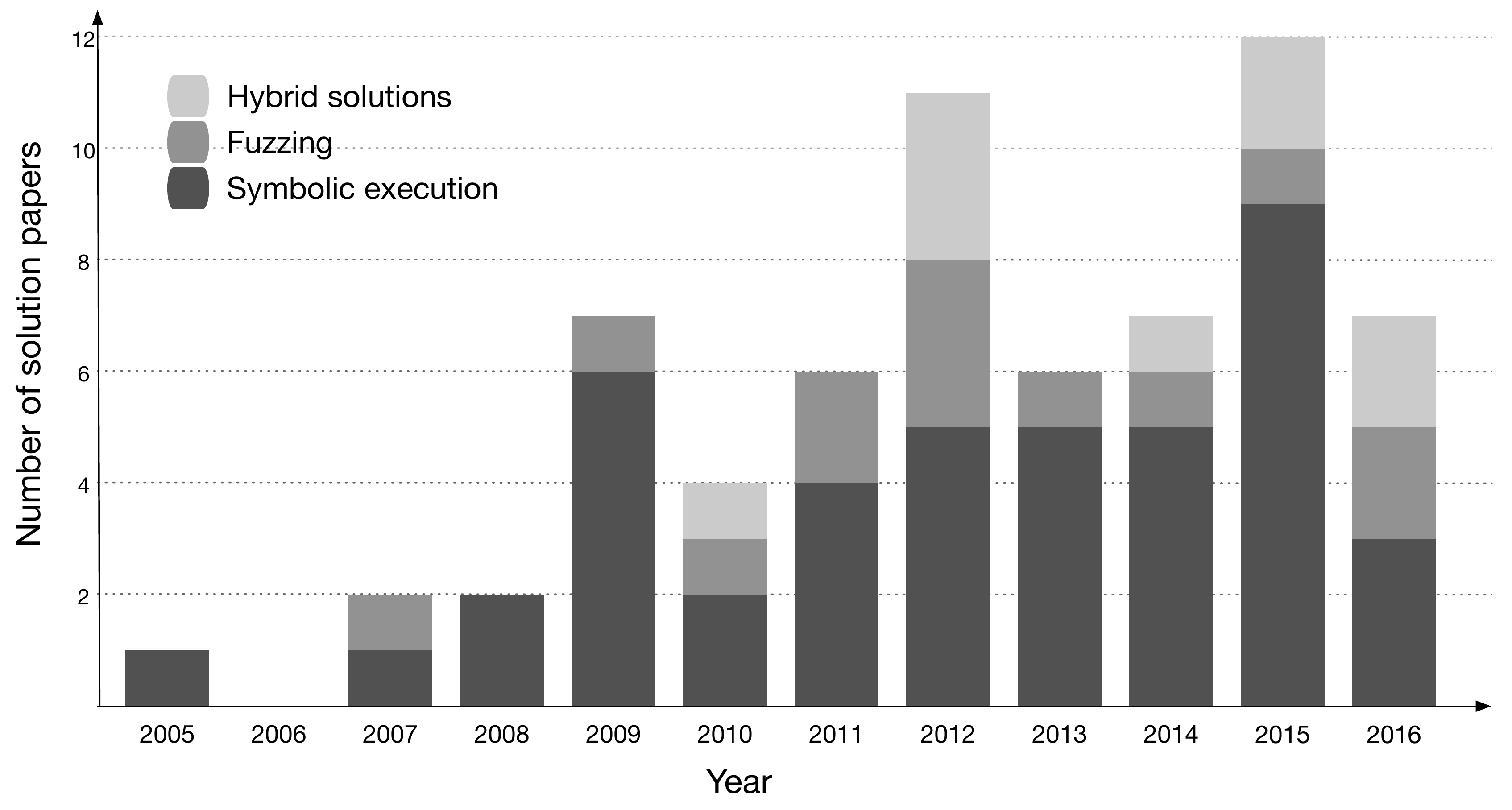}
    \caption{Number of solution proposals by year -- Vertically stacked values}
    \label{fig:solutions-trend}
\end{figure}

To answer this question we may look at the classification criteria \emph{2}. We can see from \cref{fig:solutions-trend} that most of the novel solutions proposals deal with symbolic execution, with only a few solutions proposed based on fuzzing. Possibly, due to an increase in the efficiency of constraint solvers during this period, symbolic execution became more viable as a testing technique. 
\begin{table}[]
    \centering
    \fontsize{8}{10}\selectfont
    \caption{List of all hybrid solution proposals}
    \label{tab:hybrid-papers}
    \begin{tabu}{| l | X | r |}
        \hline
        \textbf{Year} & \textbf{Title} & \textbf{Authors} \\ \hline
        2010 & \emph{Taintscope: A checksum-aware directed fuzzing tool for automatic software vulnerability detection} \cite{wang2010taintscope} & \citeauthor{wang2010taintscope} \\ \hline
        2012 & \emph{Using concolic testing to refine vulnerability profiles in Fuzzbuster} \cite{musliner2012using} & \citeauthor{musliner2012using} \\ \hline
        2012 & \emph{A directed fuzzing based on the dynamic symbolic execution and extended program behaviour model} \cite{chen2012directed} & \citeauthor{chen2012directed} \\ \hline
        2012 & \emph{Hybrid fuzz testing: Discovering software bugs via fuzzing and symbolic execution} \cite{pak2012hybrid} & \citeauthor{pak2012hybrid} \\ \hline
        2014 & \emph{Automatic software vulnerability detection based on guided deep fuzzing} \cite{cai2014automatic} & \citeauthor{cai2014automatic} \\ \hline
        2015 & \emph{Binary-oriented hybrid fuzz testing} \cite{fangquan2015binary} & \citeauthor{fangquan2015binary} \\ \hline
        2015 & \emph{Program-adaptive mutational fuzzing} \cite{cha2015program} & \citeauthor{cha2015program} \\ \hline
        2016 & \emph{Deepfuzz: Triggering vulnerabilities deeply hidden in binaries} \cite{boettinger2016deepfuzz} & \citeauthor{boettinger2016deepfuzz} \\ \hline
        2016 & \emph{Driller: Augmenting fuzzing through selective symbolic execution} \cite{stephens2016driller} & \citeauthor{stephens2016driller} \\
        \hline
    \end{tabu}
\end{table}

Our analysis found only 9 hybrid solution proposals (\cref{tab:hybrid-papers}). We identified them by looking at their texts in detail because we recognized that our search keywords may also return publications that are simply evaluation studies of symbolic execution compared with fuzzing or vice-versa, with one them not contributing anything to the proposed solution design. We discuss the exact nature of all the hybrid solutions in the next research question. 

\paragraph*{\textbf{Interpretation}} 
The first outcome of this result is that, while we tried our best to craft our search queries so as to retrieve them, most of the papers returned by the search engines were not, in fact, hybrid solutions. Even if we exclude state-of-the-art studies such as \cite{zhang2015survey,li2015survey}, there are many reasons why the solutions proposals were not all hybrid solutions. 

The first reason is that many papers, such as \cite{godefroid2008automated,jayaraman2009jfuzz}, use synonyms for symbolic execution, such as \emph{``whitebox fuzzing''} in particular, that get retrieved by our search term. Our intention in including the search keyword, ``\texttt{fuzzing}'', was to retrieve papers that contribute to \emph{blackbox fuzzing}, a technique that is relatively oblivious to the internals of the SUT. 
Whitebox fuzzing, on the other hand, is a practical approach for symbolic execution where a set of initial inputs are used to dynamically execute an instrumented program. This, as laid out by the original proposals such as \cite{king1976symbolic}, is simply a variant of pure symbolic execution. Therefore, with our search strings, we retrieved those papers too which, in our understanding, did not contribute to blackbox fuzzing at all. Another reason for retrieving non-hybrid solutions is that a lot of the returned papers, such as \cite{zhao2011h,mouzarani2016smart}, mention all our search terms in the abstract, but not as part of a solution but as a comparative subject for evaluation of their solution. 

Above are some explanations for why some of the papers found by us did not propose any new solutions and most of the solution papers did not propose hybrid strategies. After carefully classifying papers based on their abstracts, we found that 9 papers (\cref{tab:hybrid-papers}) actually proposed hybrid solutions involving fuzzing and symbolic execution. We will now discuss these hybrid solution proposals in more detail. 

\subsection{ RQ3: Technical Aspects Addressed in Solutions}\label{sec:result-rq3}
For all solution proposals (criterion \emph{2}), we look at the aspects of plain fuzzing (criterion \emph{4}) and symbolic execution (criterion \emph{5}) that have been addressed or utilized in the proposals. 
\begin{figure}[h]
    \centering
    \includegraphics[height=0.8\linewidth,keepaspectratio=true]{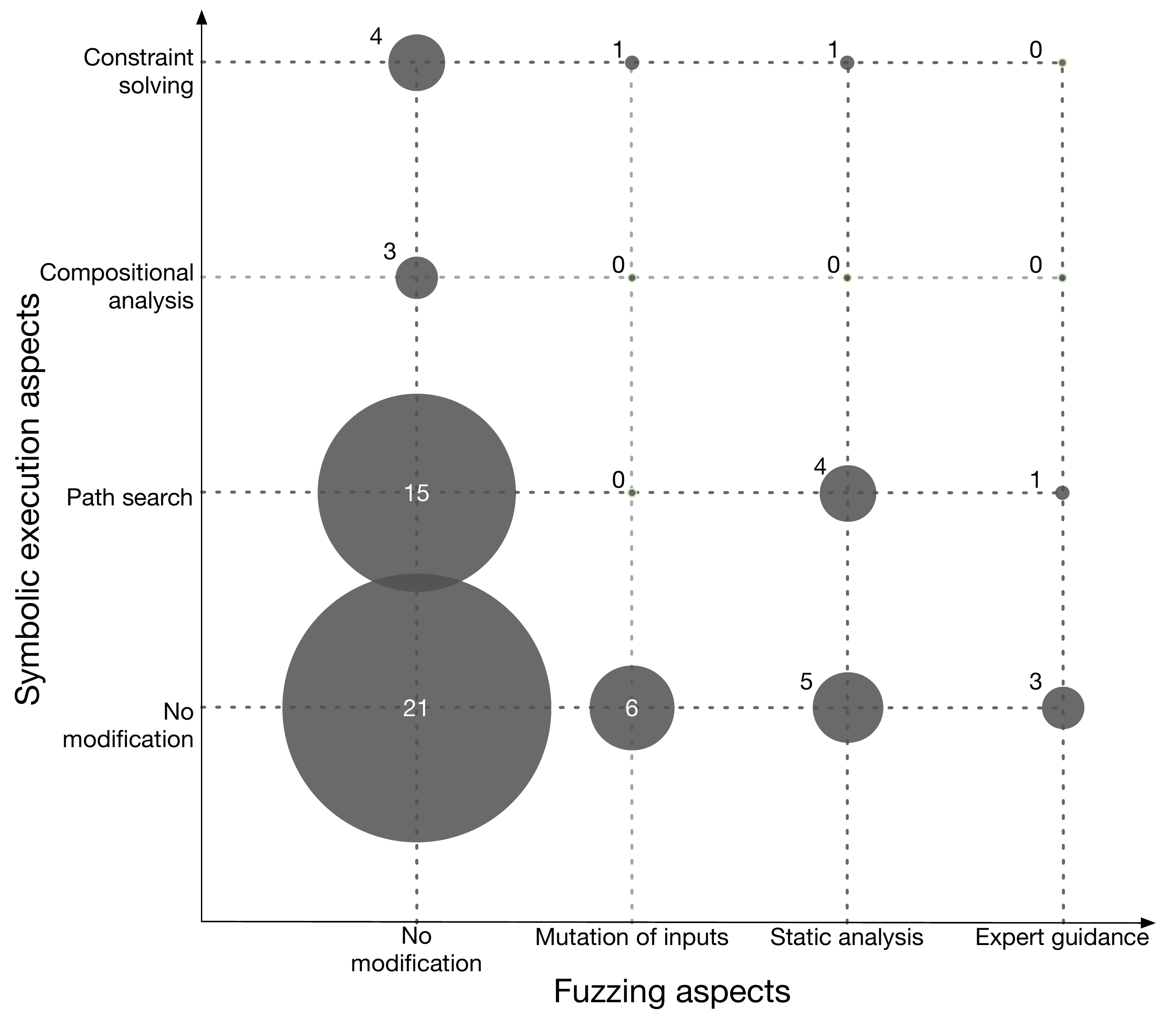}
    \caption{Technical aspects of symbolic execution and fuzzing in solution proposals}
    \label{fig:hybrid-stats}
\end{figure}
The pair-wise distribution of the technical aspects of fuzzing and symbolic execution in all solution papers is shown in \cref{fig:hybrid-stats}. The largest group of solutions (21) propose modifications to neither symbolic execution or fuzzing. Most of these contributions, such as \cite{cadar2005execution,qin2013malware}, showed up in our search results because all terms in our search string occur in them, but the proposed solutions therein use symbolic execution or fuzzing to solve domain-specific problems (such as malware detection) without adding any modifications to basic symbolic execution technique. In many cases, they may also include new target implementations, such as \cite{chen2009ewap}, of existing techniques. Other than this group, we can see that most solution proposals modify either basic symbolic execution or fuzzing technique, but not both. E.g.\ in \cite{fangquan2015binary}, the authors propose symbolic execution in binary programs when fuzzing cannot increase the coverage anymore, by focusing on uncovered paths during symbolic execution -- which means that effectively only symbolic execution technique is modified, while fuzzing is used in its original form. Similar combinations exist in techniques such as \cite{cha2015program,krishnamoorthy2011strategies}.

We notice that there are only 7 works in last 12 years that propose changes in both, symbolic execution and fuzzing. Fuzzbuster, by \cite{musliner2012using} \citeauthor{musliner2012using}, generalizes the constraints of vulnerable instructions using a modification in symbolic execution. These vulnerabilities are discovered, however, using an off-the-shelf fuzzer. This means that the tool suffers from the same drawbacks as a na\"{i}ve fuzzer, i.e.\ not enough path diversity. In \cite{chen2012directed}, \citeauthor{chen2012directed} propose a \emph{directed fuzzing} strategy that uses symbolic execution to formulate the program behavior in the form of a complex control-flow-graph. In this way, they claim that the fuzzer has an internal view of the program. Another hybrid technique, proposed by \citeauthor{pak2012hybrid} \cite{pak2012hybrid}, uses symbolic execution to gather as many unique constraints within the user-defined resource limit as possible and uses solutions to these constraints as the ``random'' input seeds for the fuzzer. Even though this technique introduces enough diversity in the seed inputs than manual entry, it relies much too heavily on the fuzzer to completely analyze all paths beyond the user-set limits. \citeauthor{cai2014automatic} \cite{cai2014automatic} introduce a tool, called \emph{Sword}, that checks the software for vulnerabilities with symbolic execution and only fuzzes those paths that are, therefore, deemed dangerous. The authors do not aim to improve the efficiency of either symbolic execution or fuzzing. In \cite{boettinger2016deepfuzz} \citeauthor{boettinger2016deepfuzz} present a probabilistic approach to treating the path-explosion problem in symbolic execution by targeting those branches in a program that are least likely to be triggered by fuzzing. In this hybrid approach, symbolic execution may also be used to solve the problem of creating input seeds for the fuzzer. In this way, the path exploration in symbolic execution is made more robust by attaching likelihoods to branches, as learned from fuzzing, and fuzzing is made more powerful by guiding the input seed generation using symbolic execution. In Driller \cite{stephens2016driller}, \citeauthor{stephens2016driller} are able to find many more vulnerabilities than na\"{i}ve symbolic execution and fuzzing tools. Their directed symbolic execution strategy is guided by a taint analysis on those inputs generated by the fuzzer that discover new paths and are, hence, deemed interesting. Driller adaptively switches between directed symbolic execution and fuzzing depending on the rate of increase in coverage of basic blocks in the program. 

In \cref{sec:result-rq2} we saw 9 papers (\cref{tab:hybrid-papers}) that had been categorized as hybrid solutions. However, as we see from \cref{fig:hybrid-stats}, there were only the 7 papers, as described above, that utilized technical aspects of, \emph{both}, fuzzing and symbolic execution in their solution. The two papers from \cref{tab:hybrid-papers} that did not get included among these 7 papers were \cite{wang2010taintscope} and \cite{fangquan2015binary}. The papers by \citeauthor{wang2010taintscope} \cite{wang2010taintscope} and \citeauthor{fangquan2015binary} \cite{fangquan2015binary} should, in our opinion, nevertheless be categorized as truly hybrid solutions. One of these papers \cite{wang2010taintscope} is a domain-specific (for applications with checksum-related operations) hybrid solution that improves fuzzing by using symbolic execution to obtain input bytes that affect checksum operations. The other \cite{fangquan2015binary} uses a simple sequence of fuzzing and symbolic execution to increase coverage in binary applications. 

\paragraph*{\textbf{Interpretation}} 
We have seen certain patterns from the results of RQ3 related to avenues used for exploiting fuzzing and symbolic execution. The most popular intersection for hybrid techniques seems to be one that includes \emph{static analysis} in fuzzing and \emph{path search} in symbolic execution. At this point, we would like to remind our readers again that the term \emph{static analysis} doesn't refer to the classical software verification and/or vulnerability discovery method, but is a general term for any helper method that helps in exposing the internal structure of the the SUT, because fuzzing itself is unaware of the system internals. Symbolic execution can be thought of as one such method. The above trend indicates that most hybrid technique proposals have focussed on solutions that deal with
\begin{enumerate*}
    \item low path-coverage in fuzzing, and
    \item path explosion problem in symbolic execution.
\end{enumerate*}
While these are indeed two critical problems in software testing, the involved technical aspects of path search or static analysis are, by no means, the only perspectives available to improve fuzzing or symbolic execution. For instance, another important bottleneck in pure symbolic execution is that associated with their underlying constraint solvers. However, only two hybrid solutions \cite{chen2012directed,pak2012hybrid} solve this technical aspect of symbolic execution using a hybrid technique. Similarly, only one work \cite{pham2016model} addresses the path-coverage problem in fuzzing by utilizing a model-based approach to overcome path-conditions in the shallow path of the program.

Therefore, based on our analysis results, we may answer RQ3 by saying that there have been some useful hybrid proposals, 7 to be exact, that have addressed or utilized technical aspects of, both, symbolic execution and fuzzing. However, there are open opportunities at various other intersections of symbolic execution and fuzzing aspects that have not been utilized yet. 
\section{Summarizing the state-of-the-art}\label{sec:interpretation}
Having seen the results of all research questions and their interpretations, we will now summarize our findings of the state-of-the-art in hybrid symbolic execution and fuzzing. 

\paragraph{\textbf{High-level implications}}
The first inference we may draw from the results is that most of the solution proposals we obtained with our systematic database search were \emph{not} truly hybrid studies. Approximately 87\% of all solution proposals were only related to either, symbolic execution or fuzzing. Even though we were only interested in mapping out hybrid solutions with symbolic execution and fuzzing, these papers ended up in our search results because their abstracts contained the search keywords, either because they compared one technique with the other, or used them as helping techniques only. However, many of these non-hybrid papers were not merely included as a side-effect. For example, the paper by \citeauthor{cha2015program} \cite{cha2015program} proposes a novel solution for optimizing seed input generation in fuzzing by tainting bits in input vectors that correspond to certain branching conditions in the program. This is an interesting use of whitebox program information to empower an input mutation strategy. However, since this technique, and others such as \cite{cui2012instruction,pham2015hercules}, do not propose improvements in or utilization of, both, symbolic execution and fuzzing, they were not considered as truly hybrid solutions. 

To properly study the truly hybrid solutions (13\% of all solution papers), it was essential to classify their symbolic execution and fuzzing technical aspects (if any), thereby creating \emph{9} slots where those techniques could effectively fit that optimize both, symbolic execution and fuzzing. As we showed in \cref{sec:result-rq3}, only 4 of these 9 open avenues for hybrid techniques were seen amongst the solution proposals. The most popular \cite{musliner2012using,cai2014automatic,boettinger2016deepfuzz,stephens2016driller} avenue among these has been \emph{static analysis+path search}. As described in \cref{sec:classification}, these hybrid solutions, generally, propose to alleviate path-explosion problem in symbolic execution by bypassing easy branching-condition checks with fuzzing. At the same time, the whitebox view obtained from symbolic execution could be used to guide the fuzzer towards more unseen paths than before. The second most popular avenue \cite{chen2012directed,pak2012hybrid} for hybrid techniques is to use fuzzing to take some load off the inefficient \emph{constraint solving} issues associated with symbolic execution. In the same papers, static analysis and mutation strategies of fuzzing are also improved using symbolic execution aspects. 

\paragraph{\textbf{Challenges and gap analysis}}
By enumerating the technical aspects of symbolic execution and fuzzing, we found that there are, indeed, many opportunities to explore w.r.t.\ more efficient techniques of automatically generating test-cases and, hence, finding vulnerabilities in systems. However, from the papers that we analyzed, some themes emerged that point to common difficulties in dealing with the low path-coverage and path-explosion problems in fuzzing and symbolic execution, respectively. Firstly, many papers \cite{stephens2016driller,fangquan2015binary,banescu2016code} demonstrated a depth-wise coverage trend with fuzzing and symbolic execution. Fuzzers, such as AFL \cite{afl}, tend to achieve low coverage in the shallow parts (closer to the program entry points) of the program. Symbolic execution tools, such as KLEE \cite{cadar2008klee}, are able to achieve high coverage in shallow depths, but not in deeper parts of the program. The challenge for researchers, then, is to find hybrid solutions involving fuzzing and symbolic execution that deal with these particular implications of both techniques. 

The second challenge for researchers is to find effective solutions for general purpose software. Only a small minority of papers such as \cite{boettinger2016deepfuzz,pak2012hybrid}, propose hybrid solutions whose evaluation results may be generalized for a large variety of programs, and only at a particular level of abstraction or choice of programming languages. Most of the papers, such as \cite{wang2010taintscope,fangquan2015binary,stephens2016driller}, present solutions to deal with problems arising due to the peculiarities of the SUTs, such as DARPA challenge candidate programs or checksum applications. These hybrid solutions, then, happen to be methods involving aspects of fuzzing and symbolic execution. This leads us to conclude that there is a severe shortage of generalizable hybrid solutions to deal with the flaws of fuzzing and symbolic execution. 

Another gap in the research which is interesting to draw attention to is the intersections involving compositional analysis with symbolic execution. The idea of compositional analysis, broadly speaking, is to isolate and treat individual components of a program with symbolic (or concolic) execution, often to find vulnerabilities in them \cite{ognawala2016macke,christakis2015ic,christakis2015proving,ma2011directed}. The compositional analysis frameworks, then, use some summarization technique to represent the vulnerable paths inside components and perform directed symbolic execution from program entry points to prove the feasibility of these vulnerabilities. The main challenge for utilizing compositional aspect for hybrid techniques is the manual effort that is typically required to generate seed inputs for fuzzing. Such an effort for individual components may make it infeasible for large programs. However, we provide some concrete ideas to solve this, in \cref{sec:visions}. 

\paragraph{\textbf{Evaluations and Benchmarks}}
One finding w.r.t.\ research in fuzzing and symbolic execution is that there is an absence of consistent evaluation criteria and benchmarks. First, many solution proposals are evaluated on limited datasets that do not generalize over application types. Parsers for different files and data types, in particular, are popular candidates for evaluation. For instance, \citeauthor{godefroid2008automated} \cite{godefroid2008automated} and Christakis et al.\ \cite{christakis2015proving} use almost the same dataset of Windows-based ANI parsers for evaluating their proposals. 

Secondly, there are various examples \cite{chen2013design,godefroid2008grammar} like the where evaluated programs are not freely available and, hence, comparison with other competing techniques is not usually possible. This also includes there are those works \cite{wang2010taintscope,zhao2011h,stephens2016driller} that evaluate their programs on very limited or non-free datasets that may suggest over-fitting of their techniques to that set alone, without scope for generalization. 

Thirdly, in several other works, such as \cite{pham2015hercules} and \cite{pham2016model}, the authors build on their previous works by including the same applications for evaluation as their past papers and showing efficiency or effectiveness gains over them. It may be argued for these papers that demonstrating improvements over programs used for evaluation in other similar works by different authors may be more substantial proofs of improvements. 

Finally, there are those works \cite{wang2010taintscope,zhao2011h,stephens2016driller} that evaluate their programs on very limited or non-free datasets that may suggest over-fitting of their techniques to that set alone, without scope for generalization. 

Overall, we observed in the papers we analyzed that no two distinct set of authors have used the same set of programs (or, in most cases, even evaluation criteria) as any other set of authors for demonstrating similar effects. We believe, therefore, that there is a need for more cross-validation studies and comparative results in the fuzzing and symbolic execution related solution proposals. 

\section{Future Visions}\label{sec:visions}
In our opinion, should the open avenues be properly addressed, we would see, amongst possible others, the following hybrid technique proposals:
\begin{enumerate}[(1)]
	\item A more robust hybrid implementation that spends less analysis time inside constraint solver. Most of the existing hybrid approaches have, somewhat successfully, tried to tackle the problem of path-explosion in symbolic execution, but the constraint solver still remains a bottleneck.
    This could be solved by the intersections that, especially, optimize \emph{constraint solving} with, e.g.\ \emph{input mutation}, \emph{static analysis} or \emph{expert guidance}. These avenues are largely unexplored, as seen in \cref{sec:result-rq3}.  
	\item A more efficient fuzzer that covers more paths than existing ones. This could be achieved by the intersections that optimize \emph{input mutation} or \emph{expert guidance}, viz.\ through \emph{path search}, \emph{compositional analysis} or \emph{constraint solving}. This is also a gap in hybrid techniques' research, which may be plugged with proposals using aspects of symbolic execution, e.g.\ thresholding of SMT queries, assertion statements related to security properties, etc.
    \item A more efficient hybrid solution to find vulnerable components in large-scale software systems. There have been many contributions in targeted vulnerability detection with symbolic execution \cite{christakis2015proving,ognawala2016macke} and fuzzing \cite{shortt2015hermes}. However, as explained in \cref{sec:interpretation}, the main challenge in combining fuzzing with compositional symbolic execution is the difficulty in generating manual test-cases for individual components. However, compositional symbolic execution may be helpful in generating test-cases for components, while targeted or compositional fuzzing \cite{serebryany2016continuous,shortt2015hermes} may provide quick path-coverage inside components. 
\end{enumerate}
Generally speaking, an internal view of the system (static analysis or expert guidance) would let the fuzzer instrument program branches. This would be, then, useful for the symbolic execution engine to guide a modified path search algorithm. We may use code instrumentation and methods like targeted symbolic execution \cite{ma2011directed,christakis2015ic,ognawala2016macke} for focussing on uncovered branches. On the side of the fuzzer, symbolic execution would be an effective way to generate (diverse) seed inputs, which affects the efficiency of the fuzzer in discovering diverse paths too. One may also use the techniques described in \cite{pak2012hybrid} to threshold constraints in symbolic execution, so that constraint solvers may not get stuck in only a few paths. 
\section{Conclusion}\label{sec:conclusion}\vspace{-1ex}
In this paper, we have presented an exploration of hybrid techniques of symbolic execution and fuzzing, in terms of their technical aspects. Using a systematic approach for mapping study, we have shown that only a few possible technical aspects have been addressed in designs of hybrid techniques. Most of the hybrid technique proposals in academia have not utilized the flexibility of individual technical aspects enough, as discussed in \cref{sec:interpretation}. 

We believe that this survey effectively maps the state-of-the-art in hybrid techniques and provides ample evidence of the gaps that exist therein. We have, with examples of hybrid techniques, argued that an ideal hybrid method would alleviate the drawbacks of \emph{both}, symbolic execution (path explosion and constraint solving) and fuzzing (low coverage). All potential avenues for doing so, however, have not been addressed adequately in literature. In the future, we propose to develop and evaluate implementations of hybrid techniques, specifically at the intersections that have not been utilized so far. Such prototypes would be compared with symbolic execution and fuzzing, and their results would be interpreted in the context of specific disadvantages of the na\"{i}ve techniques. 
\bibliographystyle{plainnat}
\bibliography{literature}

\end{document}